# Markowitz Variance May Vastly Undervalue or Overestimate Portfolio Variance and Risks

Victor Olkhov

Independent, Moscow, Russia

victor.olkhov@gmail.com

ORCID: 0000-0003-0944-5113

## Abstract

We consider the investor who doesn't trade shares of his portfolio. The investor only observes the current trades made in the market with his securities to estimate the current return, variance, and risks of his unchanged portfolio. We show how the time series of consecutive trades made in the market with the securities of the portfolio can determine the time series that model the trades with the portfolio as with a single security. That establishes the equal description of the market-based variance of the securities and of the portfolio composed of these securities that account for the fluctuations of the volumes of the consecutive trades. We show that Markowitz's (1952) variance describes only the approximation when all volumes of the consecutive trades with securities are assumed constant. The market-based variance depends on the coefficient of variation of fluctuations of volumes of trades. To emphasize this dependence and to estimate possible deviation from Markowitz variance, we derive the Taylor series of the market-based variance up to the $2^{nd}$ term by the coefficient of variation, taking Markowitz variance as a zero approximation. We consider three limiting cases with low and high fluctuations of the portfolio returns, and with a zero covariance of trade values and volumes and show that the impact of the coefficient of variation of trade volume fluctuations can cause Markowitz's assessment to highly undervalue or overestimate the market-based variance of the portfolio. Incorrect assessments of the variances of securities and of the portfolio cause wrong risk estimates, disturb optimal portfolio selection, and result in unexpected losses. The major investors, portfolio managers, and developers of macroeconomic models like BlackRock, JP Morgan, and the U.S. Fed should use market-based variance to adjust their predictions to the randomness of market trades.

---

This research received no support, specific grant, or financial assistance from funding agencies in the public, commercial, or nonprofit sectors. We welcome offers of substantial support.

# 1. Introduction

The variances of market securities and the variance of the portfolio composed of these securities measure the risks investors are taking. The dependence of the portfolio variance on the variances of its securities determines the heart of the matter for optimal portfolio selection. The accurate valuation of the portfolio variance that accounts for the randomness of current financial markets is the main challenge for the investors and portfolio managers. This paper investigates only the portfolio variance and doesn't consider any portfolio selection problems. Markowitz (1952) derived the portfolio variance $\Theta_M(t,t_0)$ (1.2) as the quadratic form in variables $X_j(t_0)$ of relative amounts invested into the securities with the coefficients equal to the covariances $\theta_{jk}(t,t_0)$ (1.3) of securities returns. The mean return $R(t,t_0)$ (1.1) at current time $t$ of the portfolio that was composed at time $t_0$ in the past of $j=1,..J$ securities, takes the form:

$$R(t, t_0) = \sum_{j=1}^{J} R_j(t, t_0) X_j(t_0) \qquad (1.1)$$

The functions $R_j(t,t_0)$ denote the average returns of the security $j$ at current time $t$ with respect to time $t_0$ in the past. The variables $X_j(t_0)$ in (1.1) denote the relative amounts invested into security $j$ at time $t_0$. Markowitz described the variance $\Theta_M(t,t_0)$ (1.2) of the portfolio as:

$$\Theta_M(t, t_0) = \sum_{j,k=1}^{J} \theta_{jk}(t, t_0) X_j(t_0) X_k(t_0) \qquad (1.2)$$

The functions $\theta_{jk}(t,t_0)$ (1.3) denote the covariance between securities $j$ and $k$ of the portfolio:

$$\theta_{jk}(t, t_0) = E\big[\big(R_j(t_i, t_0) - E[R_j(t_i, t_0)]\big)\big(R_k(t_i, t_0) - E[R_k(t_i, t_0)]\big)\big] \qquad (1.3)$$

We reconsider the derivation of the portfolio variance and show that Markowitz variance $\Theta_M(t,t_0)$ (1.2) describes the approximation when all volumes of consecutive trades with all securities of the portfolio are assumed constant during the averaging interval. The time series of constant trade volumes have *zero* coefficients of variation. One should consider Markowitz variance as an approximation valid for *zero* coefficient of variation of trade volumes. We recall that the coefficient of variation equals the ratio of the standard deviation of a random variable to its average value.

However, the financial markets reveal high fluctuations in the volumes of the consecutive trades with market securities, and that results in a positive coefficient of variation. To account for the impact of fluctuations of the volumes of consecutive trades with securities, we derive the market-based portfolio variance $\Theta(t,t_0)$ and its dependence on the coefficient of variation of the volumes of consecutive trades. That defines the difference from Markowitz variance $\Theta_M(t,t_0)$ (1.2), which describes the case with zero coefficient of variation.

To evaluate the difference between the values of market-based variance $\Theta(t,t_0)$, which accounts for the random trade volumes, and Markowitz variance $\Theta_M(t,t_0)$ (1.2), which ignores them, we derive the Taylor series of the market-based variance up to the $2^{nd}$ term by the coefficient of variation of trade volumes and take Markowitz variance $\Theta_M(t,t_0)$ (1.2) as a zero approximation. To describe the market-based variance of the portfolio, we consider the investor who collected his portfolio of $j=1,...J$ securities in the past at time $t_0$, and since then has held his portfolio and not traded his shares. However, the investor is interested in current assessments of returns, variance, and risks he takes with his portfolio. To satisfy his curiosity, the investor observes the market trades with the securities of his portfolio that were made currently in the market during the averaging interval. We show how the investor should transform the time series of the observed trades that were made in the market with the securities of his portfolio to derive the time series, which model the trades with his portfolio as with a single security. The time series of trades with the portfolio describe its return and variance exactly the same as the time series of trades with any security describe its return and variance. We highlight that we derive the equal description of return and variance of any market security and of the portfolio. The Taylor series equally describe the dependence of the market-based variance of the portfolio and of each market security on their coefficients of variation of their trade volumes.

We show how the dependence of the time series of trades with the portfolio on the time series of trades with its securities determine the decomposition of the portfolio variance by its securities. We show that if one assumes that the volumes of the consecutive trades with all securities are constant during the averaging interval, the decomposition of the portfolio variance $\Theta(t,t_0)$ coincides with Markowitz's expression $\Theta_M(t,t_0)$ (1.2).

While the portfolio theory has been developed a lot since 1952 (Markowitz, 1991; Rubinstein, 2002; Boyd et al., 2024), the expression of the portfolio variance $\Theta_M(t,t_0)$ (1.2) remains the same. We restudy only this particular issue and don't consider no problems for optimal portfolio selection. The only needed reference is Markowitz's (1952) famous study.

However, we should mention studies (Tauchen and Pitts, 1983; Karpoff, 1986; 1987; Lo and Wang, 2001; Goyenko et al., 2024) and references therein that investigate the price-volume relation. We underline that our description of the market-based variance that accounts for the impact of fluctuations of the volumes of consecutive trades has nothing in common with the above models. We highlight this to avoid possible hasty, and wrong comments that our research has its roots in these studies. That is not so.

In Section 2, we show how one should transform the time series of trades with the securities of the portfolio to derive the time series, which model the trades with the portfolio as with a single security. In Section 3, we discuss Taylor series of market-based variance. In Section 4, we discuss three cases that display when Markowitz's expression $\Theta_M(t,t_0)$ (1.2) can undervalue or overestimate the market-based variance $\Theta(t,t_0)$ that dependend on the variance on the coefficient of variation of the volumes of consecutive trades. The major investors, portfolio managers, and the developers of macroeconomic models, such as BlackRock, JPMorgan, and the U.S Fed., should keep that in mind. The Conclusion is in Section 5. App. A gives a brief derivation of market-based variance. In App. B, we derive the Taylor series of the portfolio variance and take Markowitz variance $\Theta_M(t,t_0)$ (1.2) as a zero approximation. We consider three cases that illustrate when Markowitz expression $\Theta_M(t,t_0)$ (1.2) may overvalue or underestimate the market-based variance $\Theta(t,t_0)$. In App. C we derive the Taylor series of the decomposition of the portfolio variance by its securities.

## 2. How can market trades with securities model trades with portfolio?

The time series of market trades with a security determine its return and variance. However, it is common to believe that the investor who holds his portfolio of *j=1,...J* securities and doesn't trade his shares should know the returns and covariances of returns of all securities of his portfolio to assess the return and the variance of his portfolio. We show that the investor can avoid that and directly calculate the return and the variance of his portfolio using the time series that model the current trades with his portfolio as with a single market security.

*2.1 Time series of trades with securities*

We consider the investor who at time $t_0$ in the past collected his portfolio of *j=1,...J* securities and since then doesn't trade his shares. We assume that at time $t_0$, each security *j* of the portfolio had $U_j(t_0)$ shares of the value $C_j(t_0)$ and price $p_j(t_0)$ that obey trivial equations:

$$C_j(t_0) = p_j(t_0)U_j(t_0) \quad ; \quad j = 1, ...J \qquad (2.1)$$

All prices are adjusted to current time *t*. We denote $W_\Sigma(t_0)$ (2.2) the total number of all shares of all securities in the portfolio with the total value $Q_\Sigma(t_0)$ (2.2) of the portfolio at time $t_0$.

$$Q_\Sigma(t_0) = \sum_{j=1}^{J} C_j(t_0) \quad ; \quad W_\Sigma(t_0) = \sum_{j=1}^{J} U_j(t_0) \qquad (2.2)$$

The total value $Q_\Sigma(t_0)$ and the total number of shares $W_\Sigma(t_0)$ of the portfolio determine the price $s(t_0)$ (2.3) per share of the portfolio at $t_0$. From (2.2; 2.3), obtain:

$$Q_\Sigma(t_0) = s(t_0)W_\Sigma(t_0) \quad ; \quad s(t_0) = \sum_{j=1}^{J} p_j(t_0) x_j(t_0) \quad ; \quad x_j(t_0) = \frac{U_j(t_0)}{W_\Sigma(t_0)} \qquad (2.3)$$

The functions $x_j(t_0)$ denote the relative numbers of the shares $U_j(t_0)$ of security *j* in the total number of shares $W_\Sigma(t_0)$ of the portfolio.

The investor at the current time *t* observes the time series of trades that were made in the market with the securities of his portfolio during the averaging interval. For each security *j=1,...J*, we denote the trade volumes $U_j(t_i)$, values $C_j(t_i)$, and prices $p_j(t_i)$ at time $t_i$ during the averaging interval, which follow equations (2.4):

$$C_j(t_i) = p_j(t_i)U_j(t_i) \quad ; \quad j = 1,..J \qquad (2.4)$$

We recall that (2.4) describes the trades with shares of the securities $j=1,...J$ that are currently made in the market. However, the shares of the portfolio (2.1-2.3) are not traded, and their numbers in the portfolio don't change. For convenience, we assume that the trades with all securities $j$ of the portfolio are made at $t_i$ with a small constant time span $\varepsilon$ between the trades, so $t_{i+1}=t_i+\varepsilon$. If so, any averaging interval $\Delta$ (2.5) at the current time $t$ contains only a finite number $N$ of trades with each security $j$:

$$\Delta = \left[t - \frac{\Delta}{2}; t + \frac{\Delta}{2}\right] \quad ; \quad t_i \in \Delta \quad ; \quad i = 1,...N \quad ; \quad N \cdot \varepsilon = \Delta \qquad (2.5)$$

*2.2 Time series of trades with the portfolio*

The simple equations (2.4) that define the prices $p_j(t_i)$ of securities $j=1,...J$ have the important attribute: the changes of the scale $\lambda$ of the values $C_j(t_i)$ and volumes $U_j(t_i)$ of trades don't change the price $p_j(t_i)$ and its statistical properties. We use this, and for each security $j=1,2,...J$ of the portfolio, choose the scale $\lambda_j$ (2.6) that equals the ratio of the number of shares $U_j(t_0)$ (2.1) of security $j$ in the portfolio at time $t_0$ to the current total volume of trades $U_{\Sigma j}(t)$ (2.7) with security $j$ that were made in the market at current time $t$ during $\Delta$ (2.5):

$$\lambda_j = \frac{U_j(t_0)}{U_{\Sigma j}(t)} \qquad (2.6)$$

We denote the total value $C_{\Sigma j}(t)$ and the total volume $U_{\Sigma j}(t)$ (2.7) of current trades with security $j$ that were made in the market during $\Delta$ (2.5).

$$C_{\Sigma j}(t) = \sum_{i=1}^{N} C_j(t_i) = N \cdot C_j(t) \; ; \; U_{\Sigma j}(t) = \sum_{i=1}^{N} U_j(t_i) = N \cdot U_j(t) \; ; \; j = 1,..J \quad (2.7)$$

In (2.7), $C_j(t)$ denotes the current average value and the average volume $U_j(t)$ of market trades made with security $j$ during $\Delta$ (2.5). We define the normalized values $c_j(t_i)$ and volumes $u_j(t_i)$ of trades at time $t_i$ with securities $j$

$$c_j(t_i) = \lambda_j \cdot C_j(t_i) \quad ; \quad u_j(t_i) = \lambda_j \cdot U_j(t_i) \qquad (2.8)$$

The change of scales (2.8) transforms the equations (2.4) into (2.9):

$$c_j(t_i) = p_j(t_i)\, u_j(t_i) \quad or \quad \lambda_j \cdot C_j(t_i) = p_j(t_i)\, \lambda_j \cdot U_j(t_i) \qquad (2.9)$$

It is obvious that for constant $\lambda_j$ (2.6), the properties of price $p_j(t_i)$ of security $j$ that are determined by both equations (2.9) are the same. From (2.7; 2.8; 2.9), obtain that the total normalized volumes $u_{\Sigma j}(t)$ (2.10) and average normalized volumes $u_j(t)$ (2.10) of trades with each security $j$ during $\Delta$ (2.5) exactly equals its number of shares $U_j(t_0)$ (2.1) in the portfolio:

$$u_{\Sigma j}(t) = \sum_{i=1}^{N} u_j(t_i) = N \cdot u_j(t) = \frac{U_j(t_0)}{U_{\Sigma j}(t)} \sum_{i=1}^{N} U(t_i) = U_j(t_0) \quad (2.10)$$

Thus, for each security $j$, during $\Delta$ (2.5), the time series of the normalized values $c_j(t_i)$ and volumes $u_j(t_i)$ (2.8) describe the trades with the total volumes $u_{\Sigma j}(t)$ (2.10) that are precisely equal to the numbers $U_j(t_0)$ of shares of each security $j$ in the portfolio. Simply speaking, for each security $j=1,...J$, the time series of the normalized values $c_j(t_i)$ and volumes $u_j(t_i)$ (2.8) of trades during $\Delta$ (2.5) describe the trade with exactly $U_j(t_0)$ number of shares of security $j$ of the portfolio. The sum of the trades for all $j=1,...J$ securities describe the trade with exactly $W_\Sigma(t_0)$ (2.2) shares of the portfolio. Thus, the time series of the volumes $W(t_i)$ and values $Q(t_i)$ (2.11; 2.13) during $\Delta$ (2.5), exactly describe the trades with the portfolio as with a single security:

$$Q(t_i) = \sum_{j=1}^{J} c_j(t_i) \quad ; \quad W(t_i) = \sum_{j=1}^{J} u_j(t_i) \quad (2.11)$$

Alike to (2.1; 2.3; 2.4), we define the price $s(t_i)$ (2.12) of the portfolio at time $t_i$ during $\Delta$ (2.5):

$$Q(t_i) = s(t_i) W(t_i) \quad ; \quad t_i \in \Delta \quad ; \quad i = 1, ... N \quad (2.12)$$

The time series of the values $Q(t_i)$, volumes $W(t_i)$ (2.11), and prices $s(t_i)$ (2.12) describe the trades with the portfolio as with a single security (2.1-2.3). Theses time series describe the portfolio return and variance exactly the same as the time series of the volumes $U_j(t_i)$, values $C_j(t_i)$, and prices $p_j(t_i)$ of trades with each security $j=1,...J$ describe their returns and variances. We recall that the investor doesn't trade the shares of his portfolio in the market. The time series (2.11; 2.12) model the trades with the portfolio as with a single security.

*2.3 Average price and return of the portfolio as a single security*

This section describes market-based average price, return, and their variances of the portfolio as a single market security, which account for the impact of fluctuations of the volumes of consecutive trades with its securities and with the portfolio as a whole.

The total volume $W_\Sigma(t)$ (2.13) of trades with the portfolio during $\Delta$ (2.5) is constant and equals the total number of shares $W_\Sigma(t_0)$ (2.2) in the portfolio at time $t_0$.

$$W_\Sigma(t) = \sum_{i=1}^{N} W(t_i) = \sum_{j=1}^{J} \sum_{i=1}^{N} u_j(t_i) = \sum_{j=1}^{J} U_j(t_0) = W_\Sigma(t_0) \quad (2.13)$$

The current total value $Q_\Sigma(t)$ (2.14) of trades with the portfolio during $\Delta$ (2.5) depends on current prices $s(t_i)$ (2.12) and defines the average price of the portfolio $s(t)$:

$$Q_\Sigma(t) = s(t)W_\Sigma(t) = \sum_{i=1}^{N} Q(t_i) = \sum_{j=1}^{J} \sum_{i=1}^{N} c_j(t_i) = \sum_{j=1}^{J} \sum_{i=1}^{N} p_j(t_i)u_j(t_i) \quad (2.14)$$

The average price $s(t)$ (2.14; 2.15) of the portfolio and the average prices $p_j(t)$ (2.16) of its securities during $\Delta$ (2.5) take the form of volume weighted average price (VWAP) (Berkowitz et al., 1988; Duffie and Dworczak, 2021).

$$s(t) = \frac{Q_\Sigma(t)}{W_\Sigma(t)} = \frac{1}{W_\Sigma(t_0)} \sum_{i=1}^{N} s(t_i) W(t_i) \tag{2.15}$$

$$p_j(t) = \frac{c_{\Sigma j}(t)}{u_{\Sigma j}(t)} = \frac{1}{U_{\Sigma j}(t_0)} \sum_{i=1}^{N} p_j(t_i) u_j(t_i) = \frac{1}{U_{\Sigma j}(t)} \sum_{i=1}^{N} p_j(t_i) U_j(t_i) \tag{2.16}$$

From (2.14-2.16), obtain the decomposition of VWAP *s(t)* of the portfolio by VWAP prices $p_j(t)$ of its securities that is similar to the decomposition of the portfolio price *s(t₀)* (2.3) by the initial prices $p_j(t_0)$ of securities at time $t_0$ in the past:

$$s(t) = \sum_{j=1}^{J} p_j(t) \, x_j(t_0) \tag{2.17}$$

We recall that the investor doesn't trade his shares. The investor uses the current time series of trades made with securities of his portfolio in the market during *Δ* (2.5) to assess the current average price *s(t)* (2.14; 2.15) of shares of his portfolio, return, and their variances. VWAP *s(t)* (2.14; 2.15; 2.17) determines the estimate of the current return *R(t,t₀)* of the portfolio during *Δ* (2.5) with respect to its price *s(t₀)* (2.3) at time $t_0$ in the same form as the time series of trades with securities define their returns. At time $t_i$ we determine the instant return $R_j(t_i,t_0)$ of security *j* and the instant return *R(t,t₀)* of the portfolio (2.18):

$$R_j(t_i, t_0) = \frac{p_j(t_i)}{p_j(t_0)} \quad ; \quad R(t_i, t_0) = \frac{s(t_i)}{s(t_0)} \tag{2.18}$$

For convenience, we use the definition of "gross return" (2.18) instead of the usual definition of return $r(t_i,t_0) = R(t_i,t_0) - 1$. Actually, the variances of both definitions are the same. The instant return $R_j(t_i,t_0)$ (2.18) of security *j* and (2.6; 2.8; 2.16), define the average return $R_j(t,t_0)$:

$$R_j(t, t_0) = \frac{p_j(t)}{p_j(t_0)} = \frac{1}{U_\Sigma(t_0)} \sum_{i=1}^{N} R_j(t_i, t_0) u_j(t_i) = \frac{1}{U_{\Sigma j}(t)} \sum_{i=1}^{N} R_j(t_i, t_0) U_j(t_i) \tag{2.19}$$

The time series $Q(t_i)$, $W(t_i)$ (2.11), and $s(t_i)$ (2.12) define the average return *R(t,t₀)* (2.20) of the portfolio in the same form as the return $R_j(t,t_0)$ (2.19) of security *j*:

$$R(t, t_0) = \frac{s(t)}{s(t_0)} = \frac{1}{W_\Sigma(t_0)} \sum_{i=1}^{N} R(t_i, t_0) W(t_i) = \sum_{j=1}^{J} R_j(t, t_0) X_j(t_0) \tag{2.20}$$

The use of (2.17) presents the decomposition of the portfolio return *R(t,t₀)* (2.20) by the returns $R_j(t,t_0)$ (2.19) of its securities in the same form as (1.1).

## 2.4 Market-based variances of the portfolio as a single security

To estimate market-based variance $\Theta_j(t,t_0)$ (2.21) of the returns of security *j* or market-based variance $\Theta(t,t_0)$ (2.21) of the return of the portfolio:

$$\Theta_j(t, t_0) = \frac{E_m\left[\left(p_j(t_i) - p_j(t)\right)^2\right]}{p_j^2(t_0)} \quad ; \quad \Theta(t, t_0) = \frac{E_m\left[\left(s(t_i) - s(t)\right)^2\right]}{s^2(t_0)} \tag{2.21}$$

the investor should calculate market-based variance $\phi_j(t)$ of prices of security *j* and the variance *Φ(t)* (2.22) of prices of the portfolio. We use $E_m[..]$ to denote market-based variance $\phi_j(t)$ (2.22) of prices and the variance $\Theta_j(t,t_0)$ (2.21) of returns of security $j$ that account for the randomness of the volumes $U_j(t_i)$ of trades with securities during $\Delta$ (2.5).

$$\phi_j(t) = E_m\left[\left(p_j(t_i) - p_j(t)\right)^2\right] \quad ; \quad \Phi(t) = E_m\left[\left(s(t_i) - s(t)\right)^2\right] \tag{2.22}$$

We present relations (2.21; 2.22) to underline once more that the time series of the values $C_j(t_i)$, volumes $U_j(t_i)$, and prices $p_j(t_i)$ of trades with securities $j=1,...J$ assess their variances exactly in the same form as the time series of the values $Q(t_i)$, volumes $W(t_i)$ (2.11), and prices $s(t_i)$ (2.12) of trades with the portfolio as with a single market security. In App A, we show that the market-based variance $\Phi(t)$ (A.10; 2.23) of prices and the variance $\Theta(t,t_0)$ (A.12; 2.24) of returns take the form:

$$\Phi(t) = \frac{\psi^2 - 2\varphi + \chi^2}{1+\chi^2} s^2(t) \tag{2.23}$$

$$\Theta(t, t_0) = \frac{\Phi(t)}{s^2(t_0)} = \frac{\psi^2 - 2\varphi + \chi^2}{1+\chi^2} R^2(t, t_0) \tag{2.24}$$

The relations (2.23; 2.24) depend on the coefficients of variation $\psi$ and $\chi$ of the values $Q(t_i)$ and volumes $W(t_i)$ of trades with the portfolio that are equal to the ratio of their standard deviations to their averages during $\Delta$ (2.5).

$$\psi^2 = \frac{\sigma_Q^2(t)}{Q^2(t;1)} \quad ; \quad \chi^2 = \frac{\sigma_W^2(t)}{W^2(t;1)} \quad ; \quad \varphi = \frac{cov\{Q(t),W(t)\}}{Q(t;1)W(t;1)} \tag{2.25}$$

The function $\varphi$ describes the ratio of the covariance of values $Q(t_i)$ and volumes $W(t_i)$ to their average values $Q(t;1)$ and $W(t;1)$ (2.28). We omit the dependence of $\psi$, $\chi$, and $\varphi$ on time $t$. In (2.25), $\sigma_Q^2$ (2.26) denotes the square of the standard deviation of the values $Q(t_i)$ of trades:

$$\sigma_Q^2 = \frac{1}{N}\sum_{i=1}^{N}\left(Q(t_i) - Q(t;1)\right)^2 = Q(t;2) - Q^2(t;1) \tag{2.26}$$

We recall that the finite number $N$ of trades during $\Delta$ (2.5) gives the approximations of the average values $Q(t_i)$ and volumes $W(t_i)$. In (2.25), $\sigma_W^2(t)$ (2.27) denotes the square of the standard deviation of the volumes $W(t_i)$ of trades.

$$\sigma_W^2(t) = \frac{1}{N}\sum_{i=1}^{N}\left(W(t_i) - W(t;1)\right)^2 = W(t;2) - W^2(t;1) \tag{2.27}$$

The relations (2.28; 2.29) for $n=1$ determine the average $Q(t;1)$, $W(t;1)$, and for $n=2$, the averages squares $Q(t;2)$, $W(t;2)$:

$$Q(t;n) = \frac{1}{N}\sum_{i=1}^{N} Q^n(t_i) \quad ; \quad Q(t;1) = \frac{Q_\Sigma(t)}{N} \tag{2.28}$$

$$W(t;n) = \frac{1}{N}\sum_{i=1}^{N} W^n(t_i) \quad ; \quad W(t;1) = \frac{W_\Sigma(t)}{N} \tag{2.29}$$

$cov\{Q(t),W(t)\}$ (2.30) – denotes the covariance between the time series of the values $Q(t_i)$ and volumes $W(t_i)$ of the consecutive trades with the portfolio during $\Delta$ (2.5):

$$cov\{Q(t), W(t)\} = \frac{1}{N}\sum_{i=1}^{N}(Q(t_i) - Q(t;1))(W(t_i) - W(t;1)) \qquad (2.30)$$

The variances $\Phi(t)$ (2.23) and $\Theta(t,t_0)$ (2.24) define the same coefficient of variation $\mu(\psi,\chi,\varphi)$:

$$\mu(\psi, \chi, \varphi) = \frac{\Phi(t)}{s^2(t)} = \frac{\Theta(t,t_0)}{R^2(t,t_0)} = \frac{\psi^2 - 2\varphi + \chi^2}{1+\chi^2} \qquad (2.31)$$

We denote the coefficient of variation as $\mu(\psi,\chi,\varphi)$ (2.31) to highlight its dependence on the coefficient of variation $\psi$ (2.25) of trade values $Q(t_i)$, on the coefficient of variation $\chi$ (2.25) of trade volumes $W(t_i)$, and on their covariance $\varphi$ (2.25). The coefficient of variation $\mu(\psi,\chi,\varphi)$ (2.31) equally measures the degree of fluctuations of prices $s(t_i)$ and returns $R(t_i,t_0)$ of the portfolio. Due to (2.31) it is the same coefficient of variation of prices and returns. The coefficient of variation $\mu(\psi,\chi,\varphi)$ (2.31) presents $\Phi(t)$ and $\Theta(t,t_0)$ (2.23; 2.24) as:

$$\Phi(t) = \mu(\psi, \chi, \varphi) \cdot s^2(t) \qquad ; \qquad \Theta(t,t_0) = \mu(\psi, \chi, \varphi) \cdot R^2(t,t_0) \qquad (2.32)$$

If all volumes of the consecutive trades $W(t_i)$ with the portfolio are constant during $\Delta$ (2.5) and $W(t_i)=W(t;1)=W$, then $\chi=0$, $\varphi=0$, and the coefficient of variation $\mu(\psi,\chi)$ (2.31) takes the usual and simple form (2.33). The function $\sigma_s^2(t)$ denotes the square of standard deviation of prices:

$$\mu(\psi, 0, 0) = \psi_0^2 = \frac{\sigma_Q^2(t)}{Q^2(t;1)} = \frac{\sigma_s^2(t)}{s^2(t)} \qquad ; \qquad \sigma_s^2(t) = \frac{1}{N}\sum_{i=1}^{N}\bigl(s(t_i) - s(t)\bigr)^2 \qquad (2.33)$$

We denote $\psi_0$ the coefficient of variation (2.25) if one assumes that the volumes $W(t_i)$ are constant and their coefficient of variation $\chi=0$. The coefficient of variation $\mu(\psi,\chi,\varphi)$ (2.31) and hence the variances $\Phi(t)$ and $\Theta(t,t_0)$ (2.32) depend on the coefficients of variation $\chi$ of random volumes $W(t_i)$ of the consecutive trades. To clarify the dependence on the coefficients of variation $\chi$ and to study the possible deviations of the the variances $\Phi(t)$ and $\Theta(t,t_0)$ (2.32) due to fluctuations of the volumes $W(t_i)$ of consecutive trades, we derive the Taylor expansion of the coefficient of variation $\mu(\psi,\chi,\varphi)$ (2.31), and hence of the portfolio variances $\Phi(t)$ and $\Theta(t,t_0)$ (2.32) by the coefficient of variation $\chi$ up to the 2$^{nd}$ degree terms.

## 3. Taylor series of market-based variance

In this section we consider the Taylor expansions of the variance $\Theta(t,t_0)$ (2.24; 2.32) of the portfolio by the coefficient of variation $\chi$ (2.25) up to the 2$^{nd}$ degree terms. The derivation is given in App. B. We derive the Taylor series of the decompositions of the variance $\Theta(t,t_0)$ (2.24) by securities of the portfolio in App. C.

The coefficient of variation $\chi$ (2.25) of the volumes $W(t_i)$ of consecutive trades with the portfolio is positive and less than or equal to the unit. The growth of $\chi$ from 0 to 1 describes the rise of the fluctuations of the volumes $W(t_i)$ of trades with the portfolio.

## 3.1 Taylor series of the variances $\Theta(t,t_0)$ of returns

In App. B we show that the Taylor series of the variance $\Phi(t)$ (2.23) of prices and of the of the variance $\Theta(t,t_0)$ (2.24; 2.32) of returns of the portfolio have the same form (B.11; 3.1):

$$\Theta(t, t_0) = [\psi_0^2 - 2\, a \cdot \psi_0 \cdot \chi + (1 - \psi_0^2) \cdot \chi^2] \cdot R^2(t, t_0) \qquad (3.1)$$

In (2.31) we introduce the coefficient of variance $\mu(\psi,\chi,\varphi)$ and (B.2) defines $\mu(\psi,0,0)$ in case that the coefficient of variance $\chi=0$, and hence the covariance $\varphi=0$ (B.5)

$$\mu(\psi,0,0) = \psi_0^2 = \frac{\sigma_Q^2(t)}{Q^2(t;1)} = \frac{\sigma_s^2(t)}{s^2(t)} \quad ; \quad \sigma_s^2(t) = \frac{1}{N}\sum_{i=1}^{N}(s(t_i) - s(t))^2 \qquad (3.2)$$

We use the Cauchy-Schwarz-Bunyakovskii inequality (Shiryaev, 1999, p. 123) to show that the covariance $cov\{Q(t),W(t)\}$ (2.30) is proportional to the coefficients of variation $\psi$ and $\chi$ (2.25) with the coefficient proportionality $a$ (B.5). The use of (A.19) and transform (3.1):

$$\Theta(t, t_0) = \Theta_M(t, t_0) - 2a\, \Theta_M^{1/2}(t, t_0)\, R(t, t_0)\, \chi + [R^2(t, t_0) - \Theta_M(t, t_0)]\, \chi^2 \qquad (3.3)$$

$$\Theta_M(t, t_0) = \psi_0^2 \cdot R^2(t, t_0) = \sum_{j,k=1}^{J} \theta_{jk}(t, t_0)\, X_j(t_0) X_k(t_0) \qquad (3.4)$$

The function $\Theta_M(t,t_0)$ (1.2; A.19; 3.4) describes Markowitz's expression of the portfolio variance, which is valid if all volumes of trades with the securities of the portfolio and hence the volumes $W(t_i)$ of trades with the portfolio as with a single security are constant and $\chi=0$. One can use (3.2) and show that Markowitz's approximation of the portfolio variance $\Theta_M(t,t_0)$ when $\chi=0$, takes the form (3.5):

$$\Theta_M(t, t_0) = \frac{\sigma_s^2(t,t_0)}{s^2(t_0)} = \sigma_R^2(t, t_0) = \frac{1}{N}\sum_{i=1}^{N}(R(t_i, t_0) - R(t, t_0))^2 \qquad (3.5)$$

If one assumes that $W(t_i)=W$ – constant, the average return $R(t,t_0)$ (2.20) takes the form (3.6):

$$R(t, t_0) = \frac{s(t)}{s(t_0)} = \frac{1}{N}\sum_{i=1}^{N} R(t_i, t_0) \qquad (3.6)$$

## 4. How much can the coefficient of variation $\chi$ change the variance ?

We consider three extreme cases. *As the first one,* we consider the very high fluctuations (B.18) of the portfolio returns $R(t_i,t_0)$ during the averaging interval $\Delta$ (2.5), for which the coefficient of variation $\mu(\psi,0,0)=\psi_0\sim 1$ (2.31) is almost equals to 1. One can one can neglect *(1- $\psi_0^2$)* in (3.1; B.11) and can present Taylor series as (B.14):

$$\Theta(t, t_0) \sim [1 - 2\, a \cdot \chi]\, R^2(t, t_0) \quad ; \quad \Theta_M(t, t_0) \sim R^2(t, t_0) \qquad (3.7)$$

For this case (3.7; B.13; B.14), Markowitz's assessment of the portfolio variance $\Theta_M(t,t_0)$ almost takes its maximum value $R^2(t,t_0)$. High portfolio variance relates to high risks of the portfolio, and that should upset the investors. However, if the covariance $\varphi$ is positive and $a>0$ (B.5), the impact of the coefficient of variation $\chi$ can significantly reduce the value of the variance $\Theta(t,t_0)$ and thus reduce the risks of the portfolio (B.15; B.16):

$$\Theta(t,t_0) \sim [1 - 2\,a \cdot \chi]\,R^2(t,t_0) \to [1 - 2\,a]\,R^2(t,t_0) \text{ as } \chi \to 1 \;;\; 0 < a \leq 1/2$$

$$\Theta(t,t_0) \sim [1 - 2\,a \cdot \chi]\,R^2(t,t_0) \to 0 \text{ as } \chi \to \frac{1}{2a} \;;\; a \geq 1/2$$

Thus, in the case Markowitz's estimate of the portfolio variance may vastly overvalue the real portfolio variance and risks. The action of the coefficient of variation $\chi$ can decline the portfolio variance and risks. The investors and portfolio managers should keep this in mind.

***The opposite limiting case*** describes very low fluctuations of the returns (B.18; B.19) with the coefficient of variation $\mu(\psi,0,0) = \psi_0 << 1$ (2.31). For this case, Markowitz's assessment of the portfolio variance gives (3.8; B.1)8:

$$\Theta_M(t,t_0) \sim \psi_0^2 \cdot R^2(t,t_0) \ll R^2(t,t_0) \;;\; \psi_0^2 \ll 1 \qquad (3.8)$$

However, for any sign of the covariance $\varphi$ (B.5), if $\psi_0<<\chi$, the coefficient of variance can significantly increase the portfolio variance $\Theta(t,t_0)$ from its minimum (3.8) to (B.20):

$$\Theta(t,t_0) \sim [\psi_0^2 + \chi^2]R^2(t,t_0) \to \chi^2 R^2(t,t_0), \qquad \chi^2 \to 1 - \psi_0^2 \;;\; \psi_0 \ll \chi$$

The action of fluctuations of the volumes $W(t_i)$ (2.11) of trades with the portfolio as with a single security and their coefficient of variation $\chi$ can significantly increase the portfolio variance $\Theta(t,t_0)$ and risks. For this case Markowitz's estimate of the portfolio variance $\Theta_M(t,t_0)$ (3.8; 1.2), which neglects the effects of random volumes $W(t_i)$ of consecutive trades with the portfolio, may highly undervalue the portfolio variance and risks.

***The third case describes*** zero covariance $\varphi=0$ (B.5), between the values and volumes of trades. If the covariance $\varphi$ is zero and $a=0$ (B.5), then the variance $\Theta(t,t_0)$ (B.21; 3.9) grows up from its value $\Theta_M(t,t_0) = \psi_0^2\,R^2(t,t_0)$ to its maximum value $\Theta(t,t_0) = R^2(t,t_0)$, when $\chi=1$.

$$\Theta(t,t_0) \sim [\psi_0^2 + (1 - \psi_0^2) \cdot \chi^2]\,R^2(t,t_0) \to R^2(t,t_0) \text{ as } \chi^2 \to 1 \qquad (3.9)$$

If $\psi_0^2<<1$, then (3.9) presents the case where Markowitz's estimate of the portfolio variance $\Theta_M(t,t_0)=\psi_0^2\,R^2(t,t_0)<< \Theta(t,t_0) \to R^2(t,t_0)$ may underestimate the portfolio variance $\Theta(t,t_0)$ (3.9) that is determined by the coefficient of variation $\chi^2 \to 1$.

The above three limiting cases also describe the assessments of the variance of any market security. The assessments of the variance that use the implicit assumption that the volumes of consecutive trades with the security are constant may lead to overvalue or underestimate of the variance. Both mistakes can disturb the optimal portfolio and lead to unwanted losses.

The portfolio managers and investors should keep in mind the possible effects of the fluctuations of the volumes $W(t_i)$ of trades on the portfolio variance and adjust in time Markowitz's approximation of the variance to market-based assessments to avoid the excess losses due to possible overvaluation or underestimation of the variance and risks.

## 5. Conclusion

We show that the time series of trades made during the averaging interval can describe the current market-based assessments of the variances of securities and of the portfolios. Taylor series by the coefficient of variation $\chi$ assess the possible deviations from the zero approximation by Markowitz variance $\Theta_M(t,t_0)$ (1.2) that is valid for $\chi=0$. We consider three limiting cases with high and low fluctuations of returns of the portfolio and with zero covariance of trade values and volumes. We show that the impact of fluctuations of trade volumes may cause the significant overestimation or great underestimation of the portfolio variance by Markowitz's expression (1.2). The same is valid for the assessments of the variances of any tradable market securities. To avoid unexpected losses, the investors and portfolio managers should adjust in time Markowitz variance to market-based variance that accounts for the impact of random volumes of consecutive trades.

The explicit dependence of the market-based variance on the time series of values and volumes of trades during the averaging interval reveals the additional complexity of predicting the variances of securities or of the portfolio at the time horizon $T$. To make such a forecast for a particular security, one should predict the time series of the values and volumes of trades at horizon $T$ during the averaging interval, like $\Delta$ (2.5). To forecast the portfolio variance, one should predict at the horizon $T$ the time series of the values and volumes of trades with all securities of the portfolio during interval $\Delta$ (2.5). Such forecasts are impossible without the usage of reliable market and macroeconomic models and require huge market data and efforts of highly qualified researchers for the development of true macroeconomic models. Perhaps that can be implemented by the majors like BlackRock, JP Morgan, and the U.S. Fed. Most investors and portfolio managers will probably continue to use a zero approximation, described by Markowitz variance, which ignores the impact of trade volume fluctuations. Doing this, they should be ready for unexpected losses.

# Appendix A. Market-based variance

One can determine the price probability by the set of statistical moments (Shiryaev, 1999; Shreve, 2004). The market-based average price $E_m[s(t_i)]=s(t)$ (2.15) determines the $1^{st}$ statistical moment of price probability. We recall that $E_m[..]$ denotes market-based mathematical expectation. The $2^{nd}$ market-based statistical moment $E_m[s^2(t_i)]$ should be consistent with the average price $s(t)$ (2.15) and always define non-negative variance $\Phi(t)$:

$$\Phi(t) = E_m\left[(s(t_i) - s(t))^2\right] = E_m[s^2(t_i)] - s^2(t) \geq 0 \tag{A.1}$$

To fulfill (A.1) we define market-based variance $\Phi(t)$ (A.1) as follows:

$$\Phi(t) = \frac{1}{W_\Sigma(t;2)} \sum_{i=1}^N (s(t_i) - s(t))^2 W^2(t_i) \quad ; \quad W_\Sigma(t;2) = \sum_{i=1}^N W^2(t_i) \tag{A.2}$$

One can present (A.2) as follows:

$$\Phi(t) = \Phi(t;2) - 2\Phi(t;1)s(t) + s^2(t) \tag{A.3}$$

We use (2.12) and define the function $\Phi(t;2)$ as:

$$\Phi(t;2) = \frac{1}{W_\Sigma(t;2)} \sum_{i=1}^N s^2(t_i) W^2(t_i) = \frac{1}{W(t;2)} \frac{1}{N} \sum_{i=1}^N Q^2(t_i) = \frac{Q(t;2)}{W(t;2)} \tag{A.4}$$

$$Q(t;2) = \frac{1}{N} \sum_{i=1}^N Q^2(t_i) \quad ; \quad W(t;2) = \frac{W_\Sigma(t;2)}{N} \tag{A.5}$$

Relations (A.5) denote the average of squares of trade values $Q(t;2)$ and volumes $W(t;2)$. The use of (2.12; 2.13) allow present the function $\Phi(t;1)$ as:

$$\Phi(t;1) = \frac{1}{W_\Sigma(t;2)} \sum_{i=1}^N s(t_i) W^2(t_i) = \frac{1}{W(t;2)} \frac{1}{N} \sum_{i=1}^N Q(t_i) W(t_i) = \frac{E[Q(t_i) W(t_i)]}{W(t;2)} \tag{A.6}$$

One can use (2.25; 2.26; 2.28; 2.29) and present:

$$Q(t;2) = Q^2(t) + \sigma_Q^2(t) \quad ; \quad W(t;2) = W^2(t) + \sigma_W^2(t) = W^2(t)[1 + \chi^2] \tag{A.7}$$

$$E[Q(t_i) W(t_i)] = Q(t;1)W(t;1) + cov\{Q(t), W(t)\} \tag{A.8}$$

The substitution of (A.4-A.8) into (A.3), gives:

$$\Phi(t) = \frac{Q^2(t;1) + \sigma_Q^2(t) - 2s(t)Q(t;1)W(t;1) - 2s(t)cov\{Q(t),W(t)\} + s^2(t)W^2(t;1) + s^2(t)W^2(t;1)\chi^2}{W^2(t;1)[1+\chi^2]} \tag{A.9}$$

Due to (2.15; 2.28; 2.29):

$$Q^2(t;1) - 2s(t)Q(t;1)W(t;1) + s^2(t)W^2(t;1) = [Q(t;1) - s(t)W(t;1)]^2 = 0$$

We use of (2.25) and present the variance $\Phi(t)$ (A.9) as:

$$\Phi(t) = \frac{\sigma_Q^2(t) - 2s(t)cov\{Q(t),W(t)\} + s^2(t)W^2(t;1)\chi^2}{W^2(t;1)[1+\chi^2]} = \frac{\psi^2 - 2\varphi + \chi^2}{1+\chi^2} s^2(t) \tag{A.10}$$

From (A.1) obtain the $2^{nd}$ statistical moment $E_m[s^2(t_i)]$ of price:

$$E_m[s^2(t_i)] = \Phi(t) + s^2(t) = \left[\frac{\psi^2 - 2\varphi + \chi^2}{1+\chi^2} + 1\right] s^2(t) \tag{A.11}$$

The market-based variance $\Theta(t,t_0)$ of returns follows from (A.10):

$$\Theta(t,t_0) = \frac{\Phi(t)}{s^2(t_0)} = \frac{\psi^2 - 2\varphi + \chi^2}{1+\chi^2} R^2(t,t_0) \quad (A.12)$$

One can find the additional justifications in (Olkhov, 2022-2025).

*A.2 Decomposition of the portfolio variance $\Theta(t,t_0)$ by its securities*

In this subsection we show that if the volumes of trades with all securities of the portfolio are constant and hence the coefficient of variation of the volumes $W(t_i)$ of the trades with the portfolio equals zero: $\chi=0$, then the decomposition of the portfolio variance $\Theta(t,t_0)$ (A.12) by its securities coincides with Markowitz's expression of portfolio variance (1.2).

$$\Theta_M(t,t_0) = \psi_0^2 \cdot R^2(t,t_0) = \frac{\sigma_Q^2(t)}{Q^2(t)} \frac{s^2(t)}{s^2(t_0)} \quad ; \quad \psi_0^2 = \psi^2 \text{ if } \chi = 0 \quad (A.13)$$

Let us use (2.8; 2.9; 2.11) and present $\sigma_Q^2(t)$ (2.26) as:

$$\sigma_Q^2(t) = \frac{1}{N} \sum_{i=1}^{N} (Q(t_i) - Q(t;1))^2 = \frac{1}{N} \sum_{i=1}^{N} \sum_{j,k=1}^{J} [c_j(t_i) - c_j(t)][c_k(t_i) - c_k(t)]$$

The change of order of the sums, gives:

$$\sigma_Q^2(t) = \sum_{j,k=1}^{J} \frac{1}{N} \sum_{i=1}^{N} [c_j(t_i) - c_j(t)][c_k(t_i) - c_k(t)] = \sum_{j,k=1}^{J} cov\{c_j(t), c_k(t)\} \quad (A.14)$$

In (A.14) we denote covariance $cov\{c_j(t),c_k(t)\}$ (A.15) of between two normalize values $c_j(t)$ and $c_k(t)$ (2.6; 2.8) during $\Delta$ (2.5):

$$cov\{c_j(t), c_k(t)\} = \frac{1}{N} \sum_{i=1}^{N} [c_j(t_i) - c_j(t)][c_k(t_i) - c_k(t)] \quad (A.15)$$

From (2.6) obtain:

$$cov\{c_j(t), c_k(t)\} = \frac{1}{N} \sum_{i=1}^{N} [C_j(t_i) - C_j(t)][C_k(t_i) - C_k(t)] \frac{U_j(t_0)}{U_{\Sigma j}(t)} \frac{U_k(t_0)}{U_{\Sigma k}(t)}$$

Hence, the variance $\Theta_M(t,t_0)$ (A.13), takes the form:

$$\Theta_M(t,t_0) = \sum_{j,k=1}^{J} \frac{1}{N} \sum_{i=1}^{N} \frac{[C_j(t_i) - C_j(t)]}{Q(t;1)} \frac{[C_k(t_i) - C_k(t)]}{Q(t;1)} \frac{U_j(t_0)}{U_{\Sigma j}(t)} \frac{U_k(t_0)}{U_{\Sigma k}(t)} \frac{s^2(t)}{s^2(t_0)} \quad (A.16)$$

We recall that all trade volumes $U_j(t_i) = U_j$ with securities $j=1,...J$ of the portfolio are constant and hence trade volumes $W(t_i) = W$ with the portfolio also constant. If so,

$$C_j(t_i) = p_j(t_i) U_j \quad ; \quad Q(t;1) = s(t)W$$

$$\Theta_M(t,t_0) = \sum_{j,k=1}^{J} \frac{1}{N} \sum_{i=1}^{N} \frac{[p_j(t_i) - p_j(t)]}{s(t)W} \frac{[p_k(t_i) - p_k(t)]}{s(t)W} U_j U_k \frac{U_j(t_0)}{U_{\Sigma j}(t)} \frac{U_k(t_0)}{U_{\Sigma k}(t)} \frac{s^2(t)}{s^2(t_0)}$$

$$\Theta_M(t,t_0) = \sum_{j,k=1}^{J} \frac{1}{N} \sum_{i=1}^{N} \frac{[p_j(t_i) - p_j(t)]}{p_j(t_0)} \frac{[p_k(t_i) - p_k(t)]}{p_k(t_0)} \frac{U_j}{U_{\Sigma j}(t)} \frac{U_k}{U_{\Sigma k}(t)} \frac{p_j(t_0)U_j(t_0)}{s(t_0)W} \frac{p_k(t_0)U_k(t_0)}{s(t_0)W} \quad (A.17)$$

We use $x_j(t_0)$ (2.3) that denote the relative numbers of the shares $U_j(t_0)$ of security $j$ in the total number of shares $W_\Sigma(t_0)$ of the portfolio. In (A.18) $X_j(t_0)$ denotes the relative investment into security $j$ at time $t_0$:

$$\frac{U_j}{U_{\Sigma j}(t)} = \frac{1}{N} \quad ; \quad x_j(t_0) = \frac{U_j(t_0)}{W_\Sigma(t_0)} \quad ; \quad X_j(t_0) = \frac{p_j(t_0)}{s(t_0)} x_j(t_0) = \frac{p_j(t_0) U_j(t_0)}{s(t_0) W_\Sigma(t_0)} \quad \text{(A.18)}$$

Let us divide and multiply each term in A(1.7) by the initial price $p_j(t_0)$ (2.1) of security $j$:

$$\Theta_M(t, t_0) = \sum_{j,k=1}^{J} \frac{1}{N} \left[ \sum_{i=1}^{N} \frac{[p_j(t_i) - p_j(t)]}{p_j(t_0)} \frac{[p_k(t_i) - p_k(t)]}{p_k(t_0)} \right] \cdot \frac{p_j(t_0) U_j(t_0)}{s(t_0) W_\Sigma(t_0)} \cdot \frac{p_k(t_0) U_k(t_0)}{s(t_0) W_\Sigma(t_0)}$$

$$\theta_{jk}(t, t_0) = \frac{1}{N} \sum_{i=1}^{N} \frac{[p_j(t_i) - p_j(t)]}{p_j(t_0)} \frac{[p_k(t_i) - p_k(t)]}{p_k(t_0)} = \frac{1}{N} \sum_{i=1}^{N} \left( R_j(t_i, t_0) - R_j(t, t_0) \right) \left( R_k(t_i, t_0) - R_k(t, t_0) \right)$$

From above, obtain the decomposition of the portfolio variance $\Theta(t,t_0)$ (A.13) by its securities in the form (A.19) that coincides with Markowitz's expression $\Theta_M(t,t_0)$ (1.2):

$$\Theta_M(t, t_0) = \psi_0^2 \cdot R^2(t, t_0) = \sum_{j,k=1}^{J} \theta_{jk}(t, t_0) X_j(t_0) X_k(t_0) \quad \text{(A.19)}$$

We underline, that we derived the decomposition of the portfolio variance $\Theta(t,t_0)$ (A.19) by its securities using the relations (2.11-2.14) that determine the dependence of the time series of the values $Q(t_i)$ and volumes $W(t_i)$ (2.11-2.14) of consecutive trades with the portfolio on the time series of trades with its securities. The assumption that all volumes $U_j(t_i)=U_j$ of trades with the securities $j=1,...J$ are constant and hence the volumes $W(t_i)=W$ of trades with the portfolio also constant, results in zero value of the coefficient of $\chi=0$ of fluctuations of the volumes $W(t_i)$ of the trades with the portfolio.

Hence, Markowitz portfolio variance $\Theta_M(t,t_0)$ (1.2) describes the approximation when all trade volumes with the securities of the portfolio are assumed constant during $\Delta$ (2.5).

**Appendix B. Taylor Series of the variance**

The Taylor series of the coefficient of variation $\mu(\psi,\chi,\varphi)$ (2.31) by the coefficient of variation $\chi$ determine the Taylor series of the variances $\Phi(t)$ and $\Theta(t,t_0)$ (2.32).

**B.1** *Taylor Series of the coefficient of variation $\mu(\psi,\chi,\varphi)$ (2.31)*

$$\mu(\psi, \chi, \varphi) = \mu(\psi, 0, 0) + \frac{d\mu(\psi,\chi,\varphi)}{d\chi}\bigg|_{\chi=0} \chi + \frac{1}{2} \frac{d^2 \mu(\psi,\chi,\varphi)}{d\chi^2}\bigg|_{\chi=0} \chi^2 \quad \text{(B.1)}$$

From (2.33):

$$\mu(\psi, 0, 0) = \psi_0^2 = \frac{\sigma_Q^2(t)}{Q^2(t;1)} = \frac{\sigma_s^2(t)}{s^2(t)} \quad ; \quad \sigma_s^2(t) = \frac{1}{N} \sum_{i=1}^{N} (s(t_i) - s(t))^2 \quad \text{(B.2)}$$

The 1st derivative of the coefficient of variation $\mu(\psi,\chi,\varphi)$ by $d\chi$ at point $\chi=0$:

$$\frac{d\mu(\psi,\chi,\varphi)}{d\chi}\bigg|_{\chi=0} = -2 \frac{d\varphi}{d\chi}\bigg|_{\chi=0} \quad \text{(B.3)}$$

To calculate the derivative of $\varphi$ (2.25) in (B.3) by the coefficient of variation $\chi$, let us consider the Cauchy-Schwarz-Bunyakovskii inequality (B.4) (Shiryaev, 1999, p 123) that states:

$$|\varphi| = \left| \frac{cov\{Q(t), W(t)\}}{Q(t;1) W(t;1)} \right| \leq \psi \chi \quad \text{(B.4)}$$

The inequality (B.4) allows present the covariance $\varphi$ (2.25; B.4) as:

$$\varphi = \frac{cov\{Q(t),W(t)\}}{Q(t;1)W(t;1)} = a \cdot \psi \cdot \chi \quad ; \quad -1 \leq a \leq 1 \tag{B.5}$$

The sign of *a* in (B.5) describes the positive or negative covariance between the values $Q(t_i)$ and volumes $W(t_i)$ of market trades with the portfolio. From (B.5), obtain:

$$\frac{d\,\mu(\psi,\chi,\varphi)}{d\chi}\bigg|_{\chi=0} = -2\frac{d\,\varphi}{d\chi}\bigg|_{\chi=0} = -2\,a\cdot\psi_0 = -2\,a\cdot\frac{\sigma_s}{s} \tag{B.6}$$

The 2nd derivative of the coefficient of variation $\mu(\psi,\chi,\varphi)$ by $d\chi^2$ at point $\chi=0$:

$$\frac{d^2\,\mu(\psi,\chi,\varphi)}{d\chi^2}\bigg|_{\chi=0} = 2(1-\psi_0^2) \tag{B.7}$$

The substitution of (B.2; B.3; B.7) into (B.1) give Taylor series of the coefficient of variation $\mu(\psi,\chi,\varphi)$ by the coefficient of variation $\chi$:

$$\mu(\psi,\chi,\varphi) = \psi_0^2 - 2\,a\cdot\psi_0\cdot\chi + (1-\psi_0^2)\cdot\chi^2 \tag{B.8}$$

From (B.2), obtain:

$$0 \leq \psi_0^2 = \frac{\sigma_s^2(t)}{s^2(t)} \leq 1 \tag{B.9}$$

**B.2 *Taylor Series of the portfolio variances $\Phi(t)$ and $\Theta(t,t_0)$***

The relations (2.32) and (B.1-B.3; B.7) give Taylor series of the variances $\Phi(t)$ and $\Theta(t,t_0)$:

$$\Phi(t) = [\psi_0^2 - 2\,a\cdot\psi_0\cdot\chi + (1-\psi_0^2)\cdot\chi^2]\cdot s^2(t) \tag{B.10}$$

$$\Theta(t,t_0) = [\psi_0^2 - 2\,a\cdot\psi_0\cdot\chi + (1-\psi_0^2)\cdot\chi^2]\cdot R^2(t,t_0) \tag{B.11}$$

We highlight that if one assumes that the volumes $U_j(t_i)$ of trades with all securities *j=1,...J* of the portfolio and the volumes $W(t_i)$ of trades with the portfolio are constant and the coefficient of variation $\chi=0$, then the portfolio variances $\Phi(t)$ and $\Theta(t,t_0)$ $\Theta(t,t_0)$ (B.10; B.11):

$$\Phi(t)\big|_{\chi=0} = \psi_0^2 \cdot s^2(t) \leq s^2(t) \quad ; \quad \Theta_M(t,t_0) = \psi_0^2 \cdot R^2(t,t_0) \leq R^2(t,t_0) \tag{B.12}$$

In App. A we show (A.13 - A.19) that the decomposition of the portfolio variance $\Theta_M(t,t_0)$ (B.12) by its securities coincides with Markowitz's expression of the variance $\Theta_M(t,t_0)$ (1.2).

**B.3 How much can the coefficient of variation $\chi$ change the variance $\Theta(t,t_0)$**

Let us estimate the possible change of the value of the portfolio variance $\Theta(t,t_0)$ (B.11) from Markowitz's approximation of the portfolio variance $\Theta_M(t,t_0)$ (1.2; B.12), which describes the case with the coefficient of variation $\chi=0$. To do that we consider three extreme cases that highlight the possible significant change on the portfolio variance $\Theta(t,t_0)$ (B.11) from $\Theta_M(t,t_0)$ (B.12) due to growth of the coefficient of variation $\chi$ from 0 to 1.

*a)    Very high fluctuations of portfolio returns: $\psi_0 \sim 1$.*

Let us assume that fluctuations of the portfolio returns during $\Delta$ (2.5) are very high and their coefficient of variation $\mu(\psi,0,0) = \psi_0 \sim 1$ (2.31) is almost equals to 1.

$$\psi_0 \sim 1 \rightarrow 1 - \psi_0^2 \ll 1 \quad ; \quad \psi_0 \cdot \chi \sim \chi \quad ; \quad 0 \leq \chi \leq 1 \qquad (B.13)$$

Due to (B.13) one can neglect *(1- ψ₀²)* in (B.11) and the variance *Θ(t,t₀)* takes the form:

$$\Theta(t, t_0) \sim [1 - 2a \cdot \chi] R^2(t, t_0) \quad ; \quad \Theta_M(t, t_0) \sim R^2(t, t_0) \qquad (B.14)$$

In this case, Markowitz's approximation of the portfolio variance $\Theta_M(t,t_0) \sim R^2(t,t_0)$ (B.14) takes almost maximum value. The investors and portfolio managers consider high fluctuations of returns as extremely high risks for their portfolios.

However, if the covariance $\varphi$ is positive and *a>0* (B.5), then the action of the coefficient of variation $\chi$ can significantly decline the value of the market-based variance *Θ(t,t₀)* (B.14):

$$\Theta(t, t_0) \sim [1 - 2a \cdot \chi] R^2(t, t_0) \rightarrow [1 - 2a] R^2(t, t_0) \text{ as } \chi \rightarrow 1 \; ; \; 0 < a \leq 1/2 \quad (B.15)$$

$$\Theta(t, t_0) \sim [1 - 2a \cdot \chi] R^2(t, t_0) \rightarrow 0 \text{ as } \chi \rightarrow \frac{1}{2a} \; ; \; a \geq 1/2 \qquad (B.16)$$

Thus, if the covariance $\varphi$ (B.5) is positive, the impact of fluctuations of the volumes of the consecutive trades and positive values of the coefficient of variation $\chi$ can significantly decline the portfolio variance *Θ(t,t₀)* (B.16) and portfolio risks from its almost maximum value that is estimated by Markowitz's expression (1.2) to almost a zero value (B.16).

In this case, Markowitz's expression $\Theta_M(t,t_0) \sim R^2(t,t_0)$ (1.2) may vastly overvalue the portfolio variance *Θ(t,t₀)* that can be much less than *Θ_M(t,t₀)* due to the impact of the coefficient of variation $\chi$ and positive covariance $\varphi$ (B.5).

***Very low fluctuations of portfolio returns: ψ₀ <<1.***

Let us assume that the fluctuations of the returns are very low and their coefficient of variation *μ(ψ,0,0)= ψ₀ <<1* (2.31):

$$\psi_0 \ll 1 \; ; \; 1 - \psi_0^2 \sim 1 \; ; \; \Theta_M(t, t_0) \sim \psi_0^2 \cdot R^2(t, t_0) \ll R^2(t, t_0) \qquad (B.18)$$

Then, one can approximate the portfolio variance *Θ(t,t₀)* (B.11) as:

$$\Theta(t, t_0) \sim [1 - 2a \cdot y + y^2] \psi_0^2 \cdot R^2(t, t_0) \; ; \quad y = \frac{\chi}{\psi_0} \qquad (B.19)$$

In (B.19) we introduce variable *y* as the ratio of the coefficients of variations of volume and prices. In this case Markowitz's assessment $\Theta_M(t,t_0) \sim \psi_0^2 R^2(t,t_0) << R^2(t,t_0)$ (B.18) is much less than its maximum value. The investors and the portfolio managers probably may be happy and consider that the risks of their portfolios are very low.

However, in this case the impact of the coefficient of variation $\chi$ can significantly increase the portfolio variance *Θ(t,t₀)* and portfolio risks. If *ψ₀<<χ<1*, then *y>>1* and one can neglect the term *2ay* for any sign of the covariance $\varphi$ (B.5), as it is small to compare with *y²*, and obtain:

$$\Theta(t, t_0) \sim [\psi_0^2 + \chi^2] \cdot R^2(t, t_0) \rightarrow \chi^2 \cdot R^2(t, t_0) \; , \; \chi^2 \rightarrow 1 - \psi_0^2 \; ; \; \psi_0 \ll \chi < 1 \qquad (B.20)$$

In this case (B.20), the impact of the coefficient of variation $\chi$ can significantly increase the portfolio variance $\Theta(t,t_0)$ and portfolio risks, for any sign of the covariance $\varphi$ (B.5) to compare with Markowitz's estimate $\Theta_M(t,t_0) \sim \psi_0^2 \cdot R^2(t,t_0) << R^2(t,t_0)$ to its maximum value $\Theta(t,t_0) \sim \chi^2 R^2(t,t_0)$, $\chi^2 \leq 1-\psi_0^2$.

Thus, if the coefficient of variation $\psi_0^2 << 1$ then Markowitz's estimate $\Theta_M(t,t_0) \sim \psi_0^2 \cdot R^2(t,t_0)$ may significantly underestimate the portfolio variance $\Theta(t,t_0)$ that is determined by the coefficient of variation $\chi$ for rather high fluctuations of the trade volumes $W(t_i)$.

**b)    Zero covariance $\varphi$, a=0 (B.5)**

If the covariance $\varphi$ is zero and $a=0$ (B.5), (B.11) takes the form:

$$\Theta(t,t_0) \sim [\psi_0^2 + (1-\psi_0^2) \cdot \chi^2(t)] R^2(t,t_0) \quad ; \quad \Theta_M(t,t_0) \sim \psi_0^2 \cdot R^2(t,t_0) \tag{B.21}$$

If $\psi_0^2 << \chi^2 < 1$, then the covariance $\Theta(t,t_0)$ (B.21) grows up with the increasing coefficient of variation $\chi^2 < 1$ from Markowitz's the approximation $\Theta_M(t,t_0) \sim \psi_0^2 R^2(t,t_0)$ that describes the case when $\chi=0$, to the

$$\Theta_M(t,t_0) \sim \psi_0^2 \cdot R^2(t,t_0) \ll \Theta(t,t_0) \sim \chi^2(t) R^2(t,t_0) \quad ; \quad \psi_0^2 \ll \chi^2(t) < 1 \tag{B.22}$$

The relations (B.22) illustrate that in this case Markowitz's assessment $\Theta_M(t,t_0)$ of the portfolio variance can highly underestimate the variance $\Theta(t,t_0)$ and risks of the portfolio that are generated by the fluctuations of the volumes $W(t_i)$ of trades with the portfolio.

We avoid study here all possible cases and leave that for the investors and portfolio managers who can process their market trade time series and high motivation for the correct assessment of the market-based variance that accounts for the randomness of trade volumes, sign of the covariance $\varphi$ (B.5) and other factors.

**Appendix C. Taylor Series of the decomposition of variances by securities**

Taylor series of the decomposition of the portfolio variance $\Theta(t,t_0)$ of return by its securities follow from (B.8; B.11). To obtain the decomposition of the portfolio variance $\Theta(t,t_0)$ (B.11) by its securities one should use (A.19) and derive the decompositions the coefficient of variation $\chi$ of trade volumes of the portfolio, and its square $\chi^2$, by the coefficients of variation $\chi_j$ (C.1) of the trade volumes $U_j(t_i)$ of the portfolio's securities $j=1,...J$. We define the coefficient of variation $\chi_j$ (C.2) of security $j$ alike to definition (2.25):

$$\chi_j^2 = \frac{\sigma_{Uj}^2(t)}{U_j^2(t)} \quad ; \quad \sigma_{Uj}^2(t) = \frac{1}{N}\sum_{i=1}^{N}[U_j(t_i) - U_j(t)]^2 \tag{C.1}$$

From the definition (2.25) of the coefficient of variation $\chi$, and (2.6; 2.9; 2.11), obtain:

$$\chi^2 = \frac{\sigma_W^2(t)}{W^2(t;1)} = \frac{1}{W^2(t;1)}\frac{1}{N}\sum_{i=1}^{N}\sum_{j,k=1}^{J}(u_j(t_i) - u_j(t))(u_k(t_i) - u_k(t)) =$$

$$= \frac{1}{W^2(t;1)}\sum_{j,k=1}^{J}\frac{1}{N}\sum_{i=1}^{N}(u_j(t_i)-u_j(t))(u_k(t_i)-u_k(t)) = \frac{1}{W^2(t;1)}\sum_{j,k=1}^{J}cov\{u_j(t),u_k(t)\} =$$

$$= \sum_{j,k=1}^{J}\frac{cov\{u_j(t),u_k(t)\}}{u_j(t)\cdot u_k(t)}\frac{u_j(t)}{W(t;1)}\frac{u_k(t)}{W(t;1)} = \sum_{j,k=1}^{J}\frac{cov\{U_j(t),U_k(t)\}}{U_j(t)\cdot U_k(t)}\frac{u_j(t)}{W(t;1)}\frac{u_k(t)}{W(t;1)}$$

From above and (2.3; 2.6; 2.7; 2.10), obtain:

$$\chi^2 = \sum_{j,k=1}^{J}\chi_{jk}\, x_j(t_0)x_k(t_0) \quad ; \quad \chi_{jk} = \frac{cov\{U_j(t),U_k(t)\}}{U_j(t)\cdot U_k(t)} \tag{C.2}$$

$$cov\{U_j(t),U_k(t)\} = \frac{1}{N}\sum_{i=1}^{N}(U_j(t_i)-U_j(t))(U_k(t_i)-U_k(t)) \tag{C.3}$$

We use Cauchy-Schwarz-Bunyakovskii inequality (Shiryaev, 1999, p 123) and present the covariances (C.3) of time series of the volumes $U_j(t_i)$ and $U_k(t_i)$ of trades with securities $j$ and $k$ normalized to their average values $U_j(t)$ and $U_k(t)$ by the coefficients of variation $\chi_j$ (C.1):

$$\chi_{jk} = \frac{cov\{U_j(t),U_k(t)\}}{U_j(t)U_k(t)} = \beta_{jk}\chi_j\cdot\chi_k \quad ; \quad -1\le \beta_{jk}\le 1 \; ; \quad \beta_{jj}=1 \tag{C.4}$$

Finally, obtain the decomposition of the square of the coefficient of variation $\chi^2$ of volumes $W(t_i)$ of trades with the portfolio by the coefficients of variation $\chi_j$ (C.1) of its securities:

$$\chi^2 = \sum_{j,k=1}^{J}\beta_{jk}\,\chi_j\cdot\chi_k\cdot x_j(t_0)\cdot x_k(t_0) \tag{C.5}$$

Taylor series of the coefficient of variation $\chi$ near $\chi_j=0$ is a simple exercise:

$$\chi = \left[\sum_{j,k=1}^{J}\beta_{jk}\,\chi_j\cdot\chi_k\cdot x_j(t_0)\cdot x_k(t_0)\right]^{1/2}$$

$$\frac{d\chi}{d\chi_j}\Big|_{\chi_j=0, j=1,..J} = \sum_{k=1}^{J}\frac{\beta_{jk}\chi_k\cdot x_j(t_0)\cdot x_k(t_0)}{[\chi_k\cdot\chi_k\cdot x_k(t_0)\cdot x_k(t_0)]^{1/2}}\Big|_{\chi_j,\chi_k=0} = x_j(t_0)\sum_{k=1}^{J}\beta_{jk}$$

We introduce coefficients $\beta_j$ (C.6) to obtain the decomposition of $\chi$ near $\chi_j=0$:

$$\chi = \sum_{j=1}^{J}\beta_j\,\chi_j\cdot x_j(t_0) \quad ; \quad \beta_j = \sum_{k=1}^{J}\beta_{jk} \tag{C.6}$$

From (2.20) obtain the decomposition of the portfolio return $R(t,t_0)$ (C.7)

$$R(t,t_0) = \sum_{j=1}^{J}R_j(t,t_0)X_j(t_0) \tag{C.7}$$

and from (3.6; 1.2; A.13-A.19) obtain Markowitz portfolio variance $\Theta_M(t,t_0)=\sigma_R^2(t,t_0)$ (C.8):

$$\Theta_M(t,t_0) = \sigma_R^2(t,t_0) = \sum_{j,k=1}^{J}\theta_{jk}(t,t_0)X_j(t_0)X_k(t_0) \tag{C.8}$$

(C.1-C.8) define Taylor series of the decomposition of the portfolio variance by $\chi_j$ (C.1).

$$\Theta(t,t_0) = \left[\psi_0^2 - 2\,a\cdot\psi_0\cdot\left(\sum_{j=1}^{J}\beta_j\,\chi_j\cdot x_j(t_0)\right) + (1-\psi_0^2)\cdot\left[\sum_{j,k=1}^{J}\beta_{jk}\chi_j\cdot\chi_k\cdot x_j(t_0)\cdot x_k(t_0)\right]\right]\cdot\sum_{j,k=1}^{J}R_j(t,t_0)R_k(t,t_0)X_j(t_0)X_k(t_0)$$